\documentstyle[prl,aps,twocolumn,graphicx]{revtex}
\begin{document}
\draft

\narrowtext

\noindent
{\bf Comment on ``Robustness of a Local Fermi Liquid against Ferromagnetism
and Phase Separation''}\\[.1ex]

In a Letter \cite{EngBed}, on which ongoing research is based \cite{EngBed2}, 
Engelbrecht and Bedell (EB) studied the properties of Fermi liquids with a 
local (i.e., {\bf k}-independent) irreducible self-energy. Two of their main 
results were that such {\em local Fermi liquids} are robust against 
ferromagnetism and phase separation. In this Comment we want to point out that
the conclusions of \cite{EngBed} are {\em not\/} generally valid for Fermi 
liquids with a local self-energy. We show that the 
conclusions of EB are {\em not\/} valid for lattice models in high spatial 
dimensions ($d\rightarrow\infty$), which are the only systems known to date 
for which the self-energy is purely local. 
We argue that the authors' statement, that 
``the dynamical mean-field theories, which become exact in infinite $d$, 
should lead to results that are compatible with ours ($\cdots$)'', 
does not hold. 

We start by recalling some of the properties of lattice Fermi systems in 
$d=\infty$. In high spatial dimensions the self-energy $\Sigma ({\bf k})$ is 
{\em local} \cite{MH1}. Thus it can be seen as a functional of the local Green
function: $\Sigma =\Sigma [G_{\mbox{\scriptsize loc}}(\omega )]$, where 
$G_{\mbox{\scriptsize loc}}(\omega )={\cal N}^{-1}\sum_{\bf k}G(k)$ and 
${\cal N}$ is the number of lattice sites. {\em However\/}, the 
{\em irreducible\/} \cite{vertex} vertex functions are {\em not local\/}. 
Their momentum dependence enters explicitly through the parameter 
$x({\bf k})=d^{-1}\sum_{i=1}^{d} \cos(k_i)$ \cite{MH0}, e.g., the three
contributions to the irreducible vertex function in second order 
in a local interaction are for $d\to\infty$  dependent on 
$x({\bf p}-{\bf p'})$ (bubble), $x({\bf p}-{\bf p'})$ (parallel interaction
 lines) or $x({\bf p}+{\bf p}'+{\bf q})$ (crossing interaction lines).

EB assume that a local self-energy 
$\Sigma =\Sigma [G_{\mbox{\scriptsize loc}}(\omega )]$ {\em implies\/} a local
irreducible vertex function: $\Gamma^{{\tiny\mbox{IR}}} =
\delta\Sigma (\omega )/\delta G_{\mbox{\scriptsize loc}}(\omega ')$
(below eq.\ (4) in \cite{EngBed}).
But even in the limit $d\to \infty$, where the assumption of a local 
self-energy is best, the locality of the self-energy does {\em not\/} imply 
the locality of the irreducible vertex function, see  above. This means that
functional derivation and limit process do not commute. By extension, an 
almost local self-energy in finite dimensions does {\em not\/}
imply an almost local irreducible vertex function.
Based on this crucial point, we argue that EB's concept of a 
strictly local Fermi liquid is not realistic for finite dimensions 
(e.g. $d=3$). 

Our second point (see also \cite{GU}) is that EB assume 
{\em isotropy} of the Fermi surface. This assumption is manifest in 
their eqs.\ (3) and (5). However, for 2- and 3-dimensional 
lattice models, like the Hubbard model, the Fermi surface deviates strongly 
from a sphere near half-filling. In $d=\infty$ the anisotropy is particularly 
drastic. One finds, e.g., that major parts of the Fermi ``sphere'' are 
chopped off for $d\to \infty$\cite{MH1} and that the average angle between the 
radial direction ${\bf p}$ and the normal direction 
${\bf v}=\mbox{\boldmath $\nabla$} \varepsilon({\bf p})$
is approximately $38.8^\circ$ \cite{ossadnik}. We conclude that eq.\ (6) in 
\cite{EngBed} is inappropriate for lattice models.

Thirdly, the argument in \cite{EngBed} suggesting robustness of the local 
Fermi liquid against phase separation is not sufficient. 
EB apply the ``Pomeranchek 
criterion'' (that the compressibility be positive) only to the Fermi liquid 
state, disregarding phases with broken symmetries. However, phase separation 
is a {\em first order} transition which cannot be determined from the local 
behavior of the compressibility, as suggested in \cite{EngBed}. Instead, the 
Pomeranchek criterion should be applied globally (to all possible phases), 
before phase separation can be excluded.

Furthermore, the robustness predicted by EB against ferromagnetism disagrees 
with the manifestation of this phase in work on $d=\infty$ Hubbard models
on certain lattices \cite{GU,MU,TP}. 
Two examples in $d=\infty$, displaying a first order instability 
towards phase separation, are the model for interacting spinless 
fermions \cite{RuudGoetz} and the Hubbard model \cite{PvD}.

We conclude that the assumptions in \cite{EngBed}, isotropy of the Fermi
surface 
and  locality of the vertex function, are not realized in $d=\infty$ and
cannot be considered realistic in $d=3$. The result in
\cite{EngBed}, that local Fermi liquids are ``robust'' against ferromagnetism 
and phase separation, is not generally valid. 

We acknowledge helpful discussions with D. Vollhardt, 
M. Jarrell and J. Freericks. The work is supported 
in part  (GU and EMH) by the DFG
% Deut\-sche For\-schungs\-ge\-mein\-schaft 
under SFB 341.\\[1ex]

\noindent
{\small P. G. J. van Dongen$^{1}$, G. S. Uhrig$^{2}$ and 
                   E. M\"{u}ller-Hartmann$^{2}$}\\
\hspace*{2mm} {\small $^{1}$Theoretische Physik III, 
                Universit\"{a}t Augsburg,}\\
\hspace*{3.5mm}         {\small  86135 Augsburg, Germany}\\
\hspace*{2mm} {\small $^{2}$Institut f\"{u}r Theoretische Physik, 
                Universit\"{a}t zu K\"{o}ln, \\
\hspace*{3.5mm}          {\small 50937 K\"{o}ln, Germany}\\ 

\noindent
\hspace*{2mm} {\small Received 25 August 1998}\\
\hspace*{2mm} {\small PACS numbers: 71.27.+a, 71.28.+d, 71.30.+h} \\[-16ex]


\begin{references}
\bibitem{EngBed} J. R. Engelbrecht and K. S. Bedell, Phys.\ Rev.\ Lett.\ 
                 {\bf 74}, 4265 (1995).
\bibitem{EngBed2} K. B. Blagoev, J. R. Engelbrecht and K. S. Bedell,
                  to be published in Phil.\ Mag.\ Lett.\ (1998).
\bibitem{MH1} E. M\"{u}ller-Hartmann, Z.\ Phys.\ B {\bf76}, 211 (1989).
\bibitem{vertex} Irreducible is used for the vertex function 
as in eq. (1) in \protect\cite{EngBed} consisting
of diagrams being 2-particle irreducible in one channel.
Only the {\em totally} irreducible vertex function (2-particle irreducible in
all channels) {\em is} momentum independent.
\bibitem{MH0} E. M\"{u}ller-Hartmann, Z.\ Phys.\ B {\bf74}, 507 (1989).
\bibitem{GU} G. S. Uhrig, Phys.\ Rev.\ Lett.\ {\bf 77}, 3629 (1996). 
\bibitem{ossadnik} P. Ossadnik, Diploma thesis Universit\"{a}t zu K\"oln 
                   (1990). 
\bibitem{MU} M. Ulmke, Eur.\ Phys.\ J. B {\bf 1}, 301 (1998).
\bibitem{TP} T. Obermeier, T. Pruschke and J. Keller, Phys.\ Rev.\ {\bf B56},
          R8479 (1997).
%\bibitem{MH2} E. M\"{u}ller-Hartmann, 
%              Int. J. Mod. Phys. B {\bf 3}, 2169 (1989).
\bibitem{RuudGoetz} G. Uhrig and R. Vlaming, Phys.\ Rev.\ Lett.\ {\bf 71},
                    271 (1993); {\it ibid.} Ann. Physik (Leipzig)
                    {\bf 4}, 778, (1995).
\bibitem{PvD} P. G. J. van Dongen, Phys.\ Rev.\ Lett.\ {\bf 74}, 182 (1995).
%\bibitem{Mark} See also N. S. Vidhyadhiraja, M. H. Hettler, M. Mukherjee, 
%               M. Jarrell, and H. R. Krishnamurthy, unpublished.
\end{references}
\end{document}